\documentclass[aps,pra,notitlepage,reprint,floatfix,amsmath,amssymb]{revtex4-1}

 \usepackage[english]{babel}
\usepackage{siunitx}
\usepackage{xcolor}
\usepackage{graphicx}
\graphicspath{{figures/}}         

\usepackage[caption=false]{subfig} 
\usepackage{hyperref}
\usepackage{bbm}
\usepackage{mathrsfs}

\newcommand{\Vie}[3]{\mathop{\mathbf{#1}_{#2}^{#3}}}
\newcommand{\Cie}[3]{\mathop{\mathcal{#1}_{#2}^{#3}}}

\let\Re\relax\DeclareMathOperator{\Re}{\mathrm{Re}}
\let\Im\relax\DeclareMathOperator{\Im}{\mathrm{Im}}

\begin{document}


\title{Degree of polarization of light scattered from correlated surface and bulk disorders}
\author{Jean-Philippe Banon,$^{1,2,3}$ Ingve Simonsen,$^{2,4}$ and R\'{e}mi Carminati$^{1,5}$}

\address{$^1$Institut Langevin, ESPCI Paris, CNRS, PSL University, 1 rue Jussieu, 75005 Paris, France\\
$^2$Surface du Verre et Interfaces, UMR 125 CNRS/Saint-Gobain, 93303 Aubervilliers, France\\
$^3$Laboratoire Charles Fabry, Institut d'Optique Graduate School, CNRS, Universit\'e Paris-Saclay, 91127 Palaiseau, France\\
$^4$Department of Physics, NTNU -- Norwegian University of Science and Technology, NO-7491 Trondheim, Norway\\
$^5$Institut d'Optique Graduate School, Universit\'e Paris-Saclay, 91127 Palaiseau, France}

\date{\today}

\begin{abstract}
Using a single-scattering theory, we derive the expression of the degree of polarization of the light scattered from a layer exhibiting both surface and volume scattering. The expression puts forward the intimate connection between the degree of polarization and the statistical correlation between surface and volume disorders. It also permits a quantitative analysis of depolarization for uncorrelated, partially correlated and perfectly correlated disorders. We show that measuring the degree of polarization could allow one to assess the surface-volume correlation function, and that, reciprocally, the degree of polarization could be engineered by an appropriate design of the correlation function.
\end{abstract}

\maketitle

\section{Introduction}

Polarimetric measurements are key elements in the toolbox for the characterization of complex photonic structures, including thin films, metamaterials, photonic crystals, plasmonic gratings~\cite{Oates:2011,Brakstad:15,Wang:17}, or disordered materials such as colloidal suspensions~\cite{Lam:1993,Hielscher:97,Drozdowicz:2009} and rough surfaces~\cite{Ellis:02,Letnes:2012,Simonsen:2010}. Polarization analysis is also of great interest for systems displaying both surface and volume disorder~\cite{Lam:1994,Germer:97}. In this context, depolarization measurements have shown their ability to discriminate between surface and bulk scattering. The approach has been implemented on highly scattering samples~\cite{Sorrentini:09,Dupont:14,Ghabbach:14}, where multiple scattering from the bulk is the main source of depolarization.
Interestingly, depolarization can also reveal information on weakly scattering systems, where the interaction with light occurs chiefly through single scattering, and in which volume and surface disorders may contribute with similar weights. It is often assumed that single scattering does not produce depolarization, which is actually not true for systems exhibiting (at least) two types of disorders with different polarization responses~\cite{OL:Banon:2020}. Examples of such systems are clouds of particles of different species~\cite{Buckingham}, media with depolarizing dielectric heterogeneities~\cite{Ossikovski:14}, dielectric films with rough interfaces~\cite{Germer:PRL:2000}, or samples with a rough surface and volume dielectric fluctuations~\cite{Germer:SPIE:2000,Germer:SPIE:2001}. Recently, perfect depolarization has even been predicted in the single scattering regime, for a system combining uncorrelated surface and volume disorders~\cite{OL:Banon:2020}. 

An open question is whether depolarization of the light scattered by a system with surface and volume disorders can provide information on the existence of statistical correlations between the two types of disorder. The purpose of this paper is to examine this question in the case of weakly disordered samples, in which surface and volume disorders contribute through single scattering. To proceed, we establish a general relation between the degree of polarization of the scattered light and the cross-correlation function between the surface roughness and the dielectric fluctuations in the volume. Based on this relation, we address several issues, such as the conditions to get full depolarization of the incident light, or the possibility to engineer the surface-volume correlations to produce a prescribed value of the degree of polarization of the scattered light. 

The paper is organized as follows. In section~\ref{sec:scatt:sys}, we introduce the geometry and the statistical model, focusing on the description of the cross-correlation function between the surface and volume disorders. In section~\ref{sec:theory}, we summarize the scattering theory that was described initially in Ref.~\onlinecite{Banon:2020-1}, and derive the expression of the degree of polarization. Based on this expression, we examine in section~\ref{sec:depolarization} the general conditions to get depolarization of the scattered light. In section~\ref{sec:results}, we analyze the behavior of the degree of polarization for correlated surface-volume scattering. In particular, we discuss the possibility of maximizing depolarization, and of designing the surface-volume cross-correlation function to reach a prescribed form of the degree of polarization of the scattered light. Finally, we summarize the main results in section~\ref{sect:conclusion}.
\begin{figure*}[t]
\begin{center}
\includegraphics[width=0.37\textwidth, trim=0.3cm 0cm 0.3cm 0.3cm,clip]{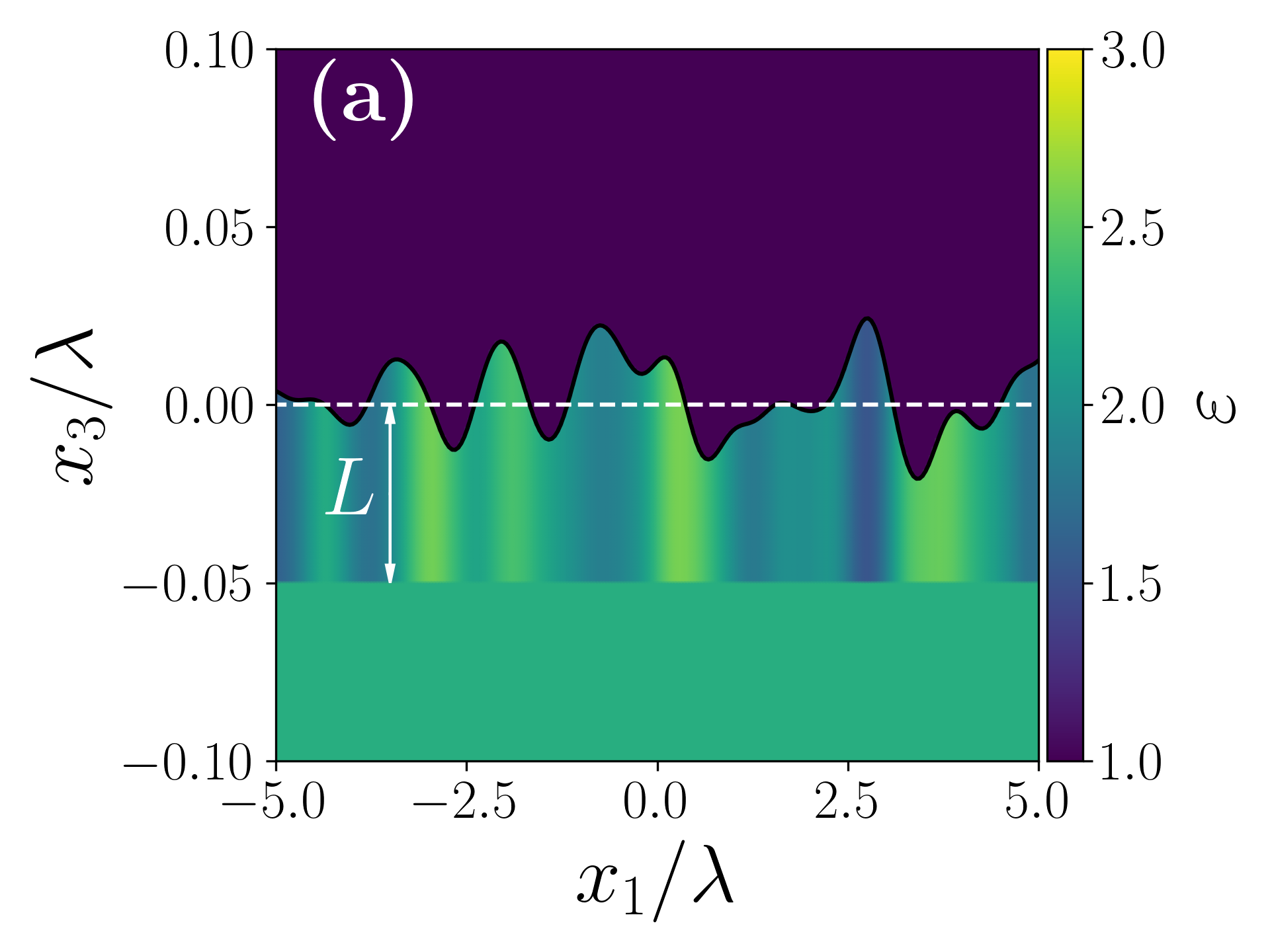}
~
\includegraphics[width=0.4\textwidth, trim=2cm 8.5cm 1.8cm 3.5cm,clip]{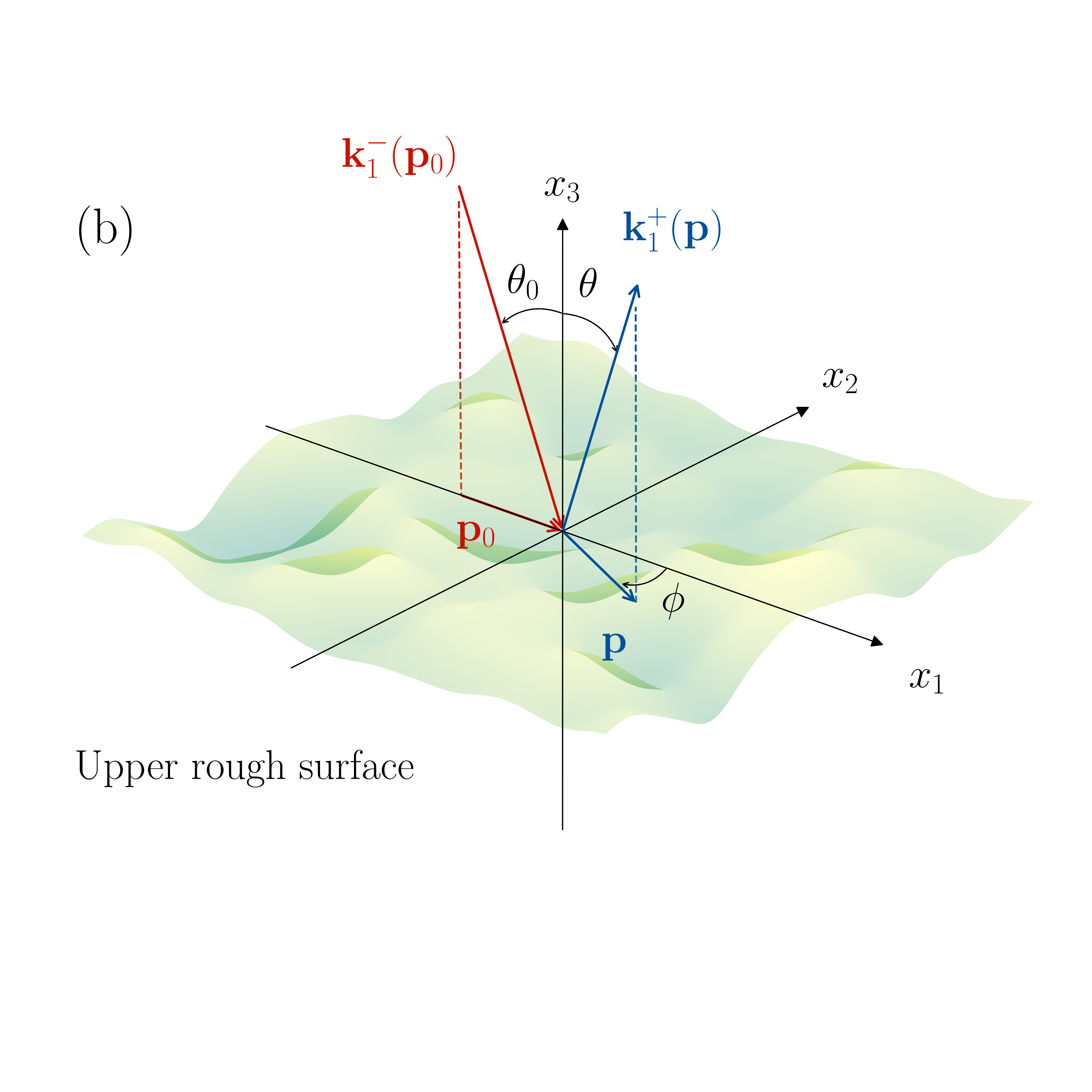}
\caption{(a) Cross-section of the scattering layer in the plane $(x_1,x_3)$, showing both a rough surface and volume dielectric fluctuations. (b) Schematics defining the incident and scattering wave vectors. The incident wave vector lies in the plane $(x_1,x_3)$, with an in-plane component $\Vie{p}{0}{}$ and a direction defined by the polar angle of incidence $\theta_0$. The scattered wave vector has an arbitrary in-plane component $\Vie{p}{}{}$, and a direction defined by the scattering angles $(\theta,\phi)$.}
\label{fig:system}
\end{center}
\end{figure*}
%

\section{Scattering geometry and statistical model}
\label{sec:scatt:sys}

We consider a scattering layer with average thickness $L$ separating two semi-infinite media, and exhibiting both surface and volume disorders~[Fig.~\ref{fig:system}(a)].  We take direction $x_3$ to be normal to the layer which is assumed to be of infinite extent along directions $x_1$ and $x_2$. The layer has a rough upper surface, described by a profile $x_3 = \zeta(\Vie{x}{\parallel}{})$, with $\Vie{x}{\parallel}{}=(x_1,x_2)$. Its lower interface is flat, and coincides with the plane $x_3=-L$. The external upper and lower media, corresponding to the regions $x_3 > \zeta(\Vie{x}{\parallel}{})$ and $x_3 < -L$, have real dielectric functions $\varepsilon_1$ and $\varepsilon_2$, respectively. The layer also exhibits volume disorder, described by a dielectric function $\varepsilon(\Vie{x}{}{})=\varepsilon_2+\Delta \varepsilon(\Vie{x}{}{})$ fluctuating around the average value $\varepsilon_2$. The geometry is depicted in Fig.~\ref{fig:system}.

In this study we will focus on the role of statistical correlations between the rough surface and the bulk dielectric fluctuations on depolarization. Depending on the dependence of $\Delta \varepsilon$ with respect to the longitudinal direction $x_3$, different types of layers can be defined.  Here we will consider dielectric fluctuations taking constant values across the layer, with $\Delta \varepsilon$ depending only on $\Vie{x}{\parallel}{}$. This type of disorder corresponds to the picture in Fig.~\ref{fig:system}(a), and was referred to as surface-like configuration in Ref.~\onlinecite{Banon:2020-1}.

In order to define the statistical model, we start by writing the dielectric function of the whole system in the form
\begin{equation}
\varepsilon (\Vie{x}{}{}) = \varepsilon_1 + \mathrm{H} \Big( \zeta(\Vie{x}{\parallel}{}) - x_3 \Big)  \Big( \varepsilon_2 -\varepsilon_1 + \Delta \varepsilon (\Vie{x}{\parallel}{}) \, \mathrm{H} (x_3 + L) \Big) \: , 
\end{equation}
where $\mathrm{H}$ is the Heaviside step function. The surface profile $\zeta$ and the dielectric fluctuation $\Delta \varepsilon$ are assumed to be realizations of correlated, zero mean and stationary Gaussian stochastic processes. In these conditions, the stochastic process defining the dielectric function $\varepsilon$ is fully characterized by $\langle \zeta (\Vie{x}{\parallel}{}) \rangle = 0$,  $\langle \Delta \varepsilon (\Vie{x}{\parallel}{}) \rangle = 0$ and 
\begin{subequations}
\begin{align}
\left\langle \zeta (\Vie{x}{\parallel}{}) \zeta (\Vie{x}{\parallel}{\prime}) \right\rangle &= \sigma_\zeta^2 \, W_\zeta (\Vie{x}{\parallel}{}-\Vie{x}{\parallel}{\prime}) \: , \label{eq:zeta:cov}\\
\left\langle \Delta \varepsilon (\Vie{x}{\parallel}{}) \Delta \varepsilon (\Vie{x}{\parallel}{\prime}) \right\rangle &= \sigma_\varepsilon^2 \,  W_\varepsilon (\Vie{x}{\parallel}{}-\Vie{x}{\parallel}{\prime}) \: , \label{eq:eps:cov}\\
\left\langle \zeta (\Vie{x}{\parallel}{}) \Delta \varepsilon (\Vie{x}{\parallel}{\prime}) \right\rangle &= \sigma_\zeta \sigma_\varepsilon \,  W_{\zeta \varepsilon} (\Vie{x}{\parallel}{}-\Vie{x}{\parallel}{\prime}) \: , \label{eq:cross}
\end{align}
\end{subequations}
where the angle brackets denote an ensemble average.  Equations~(\ref{eq:zeta:cov}) and (\ref{eq:eps:cov}) define the surface and dielectric autocorrelation functions $W_\zeta$ and $W_\varepsilon$, and standard deviations $\sigma_\zeta \geq 0$ and $\sigma_\varepsilon \geq 0$. Equation~(\ref{eq:cross}) defines the cross-correlation function of the processes $\zeta$ and $\Delta \varepsilon$.
The full definition of the stochastic process requires explicit expressions for $W_\zeta$, $W_\varepsilon$ and $W_{\zeta \varepsilon}$. A convenient model, introduced in Ref.~\onlinecite{Banon:2020-1}, assumes Gaussian autocorrelation functions given by
\begin{subequations}
\begin{align}
W_\zeta (\Vie{x}{\parallel}{}-\Vie{x}{\parallel}{\prime}) &= \exp \Bigg( - \frac{\big| \Vie{x}{\parallel}{} -\Vie{x}{\parallel}{\prime} \big|^2}{\ell_\zeta^2} \Bigg) \: , \\
W_\varepsilon (\Vie{x}{\parallel}{}-\Vie{x}{\parallel}{\prime}) &= \exp \Bigg( - \frac{\big| \Vie{x}{\parallel}{} -\Vie{x}{\parallel}{\prime} \big|^2}{\ell_\varepsilon^2} \Bigg) \: ,
\end{align}
\end{subequations}
where $\ell_\zeta$ and $\ell_\varepsilon$ are the correlation lengths of the surface roughness and the dielectric volume fluctuations, respectively. The cross-correlation function can be modeled via a power spectral density of the form
\begin{equation}
\widetilde{W}_{\zeta \varepsilon} (\Vie{p}{}{}) = \gamma(\Vie{p}{}{}) \, \widetilde{W}_\zeta^{1/2}(\Vie{p}{}{}) \widetilde{W}_\varepsilon^{1/2}(\Vie{p}{}{})  \: ,
\label{eq:cross:spectrum}
\end{equation}
where $\widetilde{f}(\Vie{p}{}{})$ denotes the two-dimensional Fourier transform of a function $f(\Vie{x}{\parallel}{})$. This specific form of the cross-spectral power density is consistent with the constraints imposed by the nature of the covariance matrix, that has to be real, symmetric and positive definite~\cite{Banon:2020-1}. The factor $\gamma (\Vie{p}{}{})$, which will be denoted by spectral correlation modulator, has to satisfy $|\gamma| \leq 1$ and $\gamma (-\Vie{p}{}{}) = \gamma^*(\Vie{p}{}{})$~\cite{Mandel:Wolf}.

\section{Degree of polarization in the single scattering regime}
\label{sec:theory}

\begin{figure}[t]
\begin{center}
\includegraphics[width=0.95\linewidth, trim=1cm 3cm 1cm 2cm,clip]{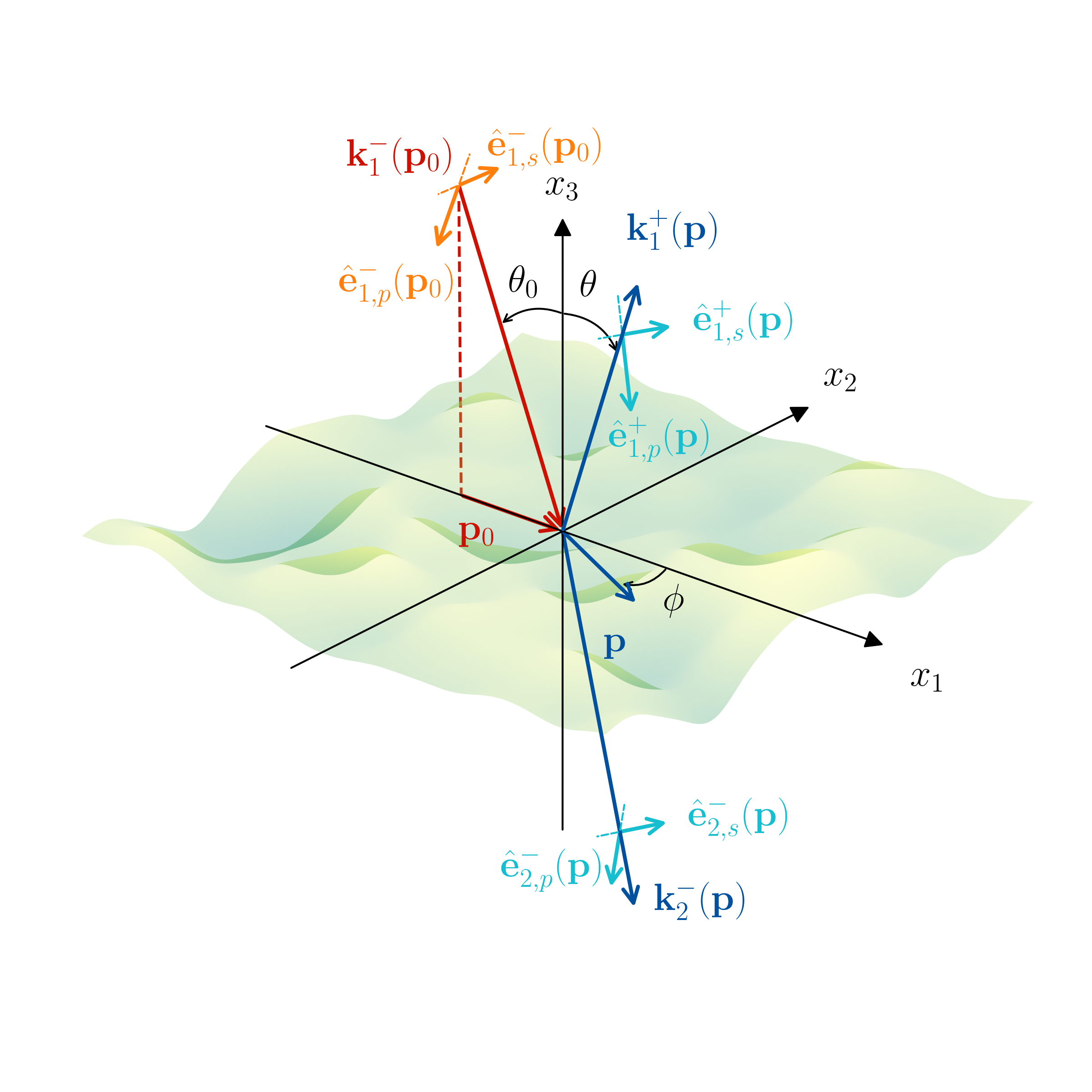}
\caption{Definition of the wave vectors for the incident and scattered fields, and of the unit vectors defining the $s$ and $p$ polarization components.}
\label{fig:system_polar}
\end{center}
\end{figure}

Our purpose is to connect the degree of polarization of the light backscattered (reflected) from the scattering layer, upon illumination by a monochromatic plane wave with angular frequency $\omega$ incident from medium 1. The complex amplitude of the incident plane wave is taken of the form
\begin{equation}
\Vie{E}{0}{} (\Vie{x}{}{}) =  \sum_{\nu = p,s }  E_{0,\nu} \, \Vie{\hat{e}}{1,\nu}{-} (\Vie{p}{0}{}) \, \exp [ i \Vie{p}{0}{} \cdot \Vie{x}{\parallel}{} - i \alpha_1(\Vie{p}{0}{}) x_3 ] \: ,
\label{eq:incident:field}
\end{equation}
where $\alpha_1(\Vie{p}{0}{})$ is the normal component of the wave vector, and $\Vie{\hat{e}}{1,\nu}{-} (\Vie{p}{0}{})$ are unit vectors defining the $s$ and $p$ polarizations. These vectors are defined in medium $j=1$ and medium $j=2$ by the following relations
\begin{subequations}
\begin{align}
&\alpha_j(\Vie{p}{}{}) = \left( \varepsilon_j k_0^2 - \Vie{p}{}{2} \right)^{1/2} \:, \: \Re (\alpha_j) \geq 0, \: \Im (\alpha_j) \geq 0 \: , \\
&\Vie{\hat{e}}{j,s}{\pm} (\Vie{p}{}{}) = \Vie{\hat{e}}{3}{} \times \Vie{\hat{p}}{}{} \: , \label{eq:wave_vector:es}\\
&\Vie{\hat{e}}{j,p}{\pm} (\Vie{p}{}{}) = \frac{\pm \alpha_j(\Vie{p}{}{}) \Vie{\hat{p}}{}{} - |\Vie{p}{}{}| \, \Vie{\hat{e}}{3}{}}{\sqrt{\varepsilon_j} k_0} \: .\label{eq:wave_vector:ep}
\end{align}
\label{eq:wave_vector}
\end{subequations}
In these relations $\Vie{\hat{p}}{}{}=\Vie{p}{}{}/|\Vie{p}{}{}|$, $\Vie{\hat{e}}{3}{}$ is the unit vector along the positive $x_3$ axis and $k_0=\omega/c=2\pi/\lambda$ with $c$ the speed of light in vacuum. The meaning of the different wave vectors and polarization vectors is illustrated in Fig.~\ref{fig:system_polar}.

The purpose of this work is to characterize the degree of polarization of the scattered field $\Vie{E}{s}{}(\Vie{x}{\parallel}{},x_3)$ for an observation point in reflection, {\it i.e.} for $x_3 > \zeta(\Vie{x}{\parallel}{})$. The Fourier transform  of the field with respect to $\Vie{x}{\parallel}{}$ can be written in the form
\begin{align}
\Vie{\widetilde{E}}{s}{} (\Vie{p}{}{},x_3) = &\sum_{\mu = p,s } \Vie{\hat{e}}{1,\mu}{+} (\Vie{p}{}{}) \sum_{\nu = p,s} R_{\mu \nu}^{}(\Vie{p}{}{},\Vie{p}{0}{}) \, E_{0,\nu}   \nonumber\\
& \times  \exp \Big( i \alpha_1(\Vie{p}{}{}) \, x_3 \Big) \: ,
\label{eq:scattered:field}
\end{align}
where the reflection amplitude $R_{\mu \nu}(\Vie{p}{}{},\Vie{p}{0}{})$ connects a scattered wave in state $(\Vie{p}{}{},\mu)$ to an incident wave in state $(\Vie{p}{0}{},\nu)$. In the single scattering regime, the scattered field can be written as the sum of a contribution from the rough surface and a contribution from the volume dielectric fluctuations~\cite{Banon:2020-1,Elson:84}. In terms of the reflection amplitude, this means that
\begin{align}
R_{\mu \nu} = R_{\zeta, \mu \nu} + R_{\varepsilon, \mu \nu} \: ,
\end{align}
where $R_{\zeta, \mu \nu}$ and $R_{\varepsilon, \mu \nu}$ are the surface and volume reflection amplitudes, respectively. For a weakly scattering layer, such that the conditions of small surface amplitude ($ \sqrt{\varepsilon_j} k_0 \sigma \ll 1$) and small thickness ($\sqrt{\varepsilon_j}  k_0 L \ll 1$) are satisfied, the reflection amplitudes have analytical expressions. They can be written as\cite{Banon:2020-1}
\begin{subequations}
\begin{align}
R_{\zeta, \mu \nu}(\Vie{p}{}{},\Vie{p}{0}{}) &= s(\Vie{p}{}{},\Vie{p}{0}{}) \, \rho_{\zeta, \mu \nu}(\Vie{p}{}{},\Vie{p}{0}{}) \: , \label{eq:Rzeta}\\
R_{\varepsilon, \mu \nu} (\Vie{p}{}{},\Vie{p}{0}{}) &= v(\Vie{p}{}{},\Vie{p}{0}{}) \: \rho_{\varepsilon, \mu \nu}(\Vie{p}{}{},\Vie{p}{0}{}) \: . \label{eq:Reps}
\end{align}
\label{eq:R}
\end{subequations}
Each reflection amplitude is the product of a random contribution from the surface or the volume and of a deterministic polarization coupling factor. The contributions from surface and volume disorders take the following forms
\begin{subequations}
\begin{align}
s(\Vie{p}{}{},\Vie{p}{0}{})  &= \frac{i k_0^2}{2 \alpha_2(\Vie{p}{}{})} \, (\varepsilon_2 -\varepsilon_1) \widetilde{\zeta}(\Vie{p}{}{}-\Vie{p}{0}{}) \: ,\\
v(\Vie{p}{}{},\Vie{p}{0}{}) &= \frac{i k_0^2}{2\alpha_2(\Vie{p}{}{})} \Delta \widetilde{\varepsilon} (\Vie{p}{}{} - \Vie{p}{0}{}) \, L \: ,
\end{align}
\label{eq:sv}
\end{subequations}
where $\widetilde{\zeta}$ and $\Delta \widetilde{\varepsilon}$ are the Fourier transforms of the surface profile function and of the dielectric fluctuation.
The polarization coupling factors are given by
\begin{widetext}
\begin{subequations}
\begin{align}
\rho_{\zeta, \mu \nu} (\Vie{p}{}{},\Vie{p}{0}{}) &= t_{12}^{(\mu)}(\Vie{p}{}{}) \, \Vie{\hat{e}}{2,\mu}{+}(\Vie{p}{}{}) \cdot \Big[ \Vie{\hat{e}}{1,\nu}{-}(\Vie{p}{0}{}) + r_{21}^{(\nu)}(\Vie{p}{0}{}) \Vie{\hat{e}}{1,\nu}{+}(\Vie{p}{0}{}) \Big] \: , \\
\rho_{\varepsilon, \mu \nu} (\Vie{p}{}{},\Vie{p}{0}{}) &= t_{12}^{(\mu)}(\Vie{p}{}{}) \, \Vie{\hat{e}}{2,\mu}{+}(\Vie{p}{}{}) \cdot \Vie{\hat{e}}{2,\nu}{-}(\Vie{p}{0}{}) \, t_{21}^{(\nu)}(\Vie{p}{0}{}) \: ,
\end{align}
\label{eq:rho}
\end{subequations}
\end{widetext}
where $r_{ji}^{(\nu)}$ and $t_{ji}^{(\nu)}$ are the Fresnel reflection and transmission amplitudes for a $\nu$-polarized plane wave incident on a planar surface from medium $i$ to medium $j$ [see for example Ref.~\onlinecite{Banon:2020-1}, Eq.~(A4)]. The polarization coupling factors depend only on the geometry of the reference system, namely, a planar interface between two homogeneous media with dielectric functions $\varepsilon_1$ and $\varepsilon_2$. Physically, they describe the polarization response of an electric dipole source radiating in the reference medium~\cite{Banon:2020-1,Banon:2019}.

The polarization coupling factors have interesting properties, that will be useful in the following. First, it can be verified that the surface and volume factors are different only for $\mu = \nu = p$. Second, it is also interesting to note that for normal incidence ($\Vie{p}{0}{} = \Vie{0}{}{}$) the two polarization coupling factors in Eq.~\eqref{eq:rho} are equal for all $\Vie{p}{}{}$ and any pair of polarization states ($\mu$, $\nu$). Finally, they are real-valued functions in the radiative region ($|\mathbf{p}| < \sqrt{\varepsilon_1} k_0$). In summary, the polarization coupling factors satisfy
\begin{subequations}
\begin{align}
&\rho_{\zeta, \mu s} (\Vie{p}{}{},\Vie{p}{0}{}) = \rho_{\varepsilon, \mu s}(\Vie{p}{}{},\Vie{p}{0}{}) \equiv \rho_{\mu s} (\Vie{p}{}{},\Vie{p}{0}{}) \label{eq:rho_mu_s} \: ,\\
&\rho_{\zeta, s p}(\Vie{p}{}{},\Vie{p}{0}{}) = \rho_{\varepsilon, s p} (\Vie{p}{}{},\Vie{p}{0}{}) \equiv \rho_{s p} (\Vie{p}{}{},\Vie{p}{0}{}) \label{eq:rho_sp} \: ,\\
&\rho_{\zeta, \mu \nu}(\Vie{p}{}{},\Vie{0}{}{}) = \rho_{\varepsilon, \mu \nu} (\Vie{p}{}{},\Vie{0}{}{})\label{eq:rho:normal} \: ,\\
&\rho_{\zeta, \mu \nu}(\Vie{p}{}{},\Vie{p}{0}{}) \:  \mathrm{and} \:  \rho_{\varepsilon, \mu \nu}(\Vie{p}{}{},\Vie{p}{0}{}) \in \mathbb{R} \: \mathrm{for} \: |\Vie{p}{}{}|, |\Vie{p}{0}{}| \leq \sqrt{\varepsilon}_1 k_0 \: . \label{eq:real}
\end{align}
\label{eq:pola:properties}
\end{subequations}

We now turn to the expression for the degree of polarization of the backscattered light. For an incident plane wave in state $(\Vie{p}{0}{},\nu)$,  the degree of polarization of a wave scattered in direction $\Vie{p}{}{}$ is defined as~\cite{born_wolf:1999}
\begin{equation}
\mathcal{P}^{(\nu)}(\Vie{p}{}{},\Vie{p}{0}{}) = \left(1 - 4 \frac{\det \Vie{J}{}{(\nu)}(\Vie{p}{}{},\Vie{p}{0}{})}{\big[ \mathrm{Tr} \Vie{J}{}{(\nu)}(\Vie{p}{}{},\Vie{p}{0}{}) \big]^2} \right)^{1/2} \: ,
\label{eq:DOP}
\end{equation}
where $\Vie{J}{}{(\nu)}$ is the Jones coherence matrix with matrix elements
\begin{equation}
J^{(\nu)}_{\mu \mu^\prime} (\Vie{p}{}{},\Vie{p}{0}{}) = \left\langle R_{\mu \nu}^{} (\Vie{p}{}{},\Vie{p}{0}{}) R_{\mu^\prime \nu}^{*}  (\Vie{p}{}{},\Vie{p}{0}{}) \right\rangle \: .
\label{eq:Jones}
\end{equation}
We see that the degree of polarization is directly obtained from the reflection amplitude $R_{\mu \nu}$. It characterizes the statistical correlation between different vector components of the scattered field, given a state of polarization of the incident field (different incident states can lead to different degrees of polarization). Explicit expressions for the determinant and the trace of the coherence matrix can be obtained by inserting Eq.~(\ref{eq:R}) into Eq.~(\ref{eq:Jones}). For {\it uncorrelated} surface and volume disorders, we would simply have
\begin{subequations}
\begin{align}
\det \Vie{J}{\mathrm{unco}}{(\nu)} = &\left\langle |s|^2 \right\rangle \left\langle |v|^2 \right\rangle \left| \rho_{\zeta, p \nu} \rho_{\varepsilon, s \nu} - \rho_{\varepsilon, p \nu} \rho_{\zeta, s \nu} \right|^2, \label{eq:detJucorr}\\
\mathrm{Tr} \Vie{J}{\mathrm{unco}}{(\nu)} = &\left\langle |s|^2 \right\rangle \left( |\rho_{\zeta, p \nu}|^2 + |\rho_{\zeta, s \nu}|^2 \right) \nonumber\\
&+ \left\langle |v|^2 \right\rangle \left( |\rho_{\varepsilon, p \nu}|^2 + |\rho_{\varepsilon, s \nu}|^2 \right) \: . \label{eq:TrJucorr}
\end{align}
\label{eq:DOP:uncorrelated}
\end{subequations}
In the presence of surface-volume correlations, additional contributions have to be taken into account, and we find that
\begin{widetext}
\begin{subequations}
\begin{align}
\det \Vie{J}{}{(\nu)} = \:  &\det \Vie{J}{\mathrm{unco}}{(\nu)} + 2 \Re \left( \left\langle s v^* \right\rangle \rho_{\zeta, s\nu} \rho_{\varepsilon, s\nu}^* \right) \Big[ \left\langle |s|^2\right\rangle |\rho_{\zeta, p \nu} |^2 + \left\langle |v|^2\right\rangle |\rho_{\varepsilon, p \nu} |^2 \Big] \nonumber\\
&+ 2 \Re \left( \left\langle s v^* \right\rangle \rho_{\zeta, p\nu} \rho_{\varepsilon, p \nu}^* \right) \Big[ \left\langle |s|^2\right\rangle |\rho_{\zeta, s \nu} |^2 + \left\langle |v|^2\right\rangle |\rho_{\varepsilon, s \nu} |^2 \Big]  \nonumber\\
&+ 4 \Re \left( \left\langle s v^* \right\rangle \rho_{\zeta, p\nu} \rho_{\varepsilon, p \nu}^* \right) \Re \left( \left\langle s v^* \right\rangle \rho_{\zeta, s\nu} \rho_{\varepsilon, s\nu}^* \right) \nonumber\\
&- \Big[ \left\langle |s|^2\right\rangle \rho_{\zeta, p \nu} \rho_{\zeta, s \nu}^* + \left\langle |v|^2\right\rangle \rho_{\varepsilon, p \nu} \rho_{\varepsilon, s \nu}^* \Big] \Big[ \left\langle s v^* \right\rangle \rho_{\zeta, s \nu} \rho_{\varepsilon, p \nu}^* + \left\langle v s^*\right\rangle \rho_{\varepsilon, s \nu} \rho_{\zeta, p \nu}^* \Big] \nonumber\\
&- \Big[ \left\langle |s|^2\right\rangle \rho_{\zeta, s \nu} \rho_{\zeta, p \nu}^* + \left\langle |v|^2\right\rangle \rho_{\varepsilon, s \nu} \rho_{\varepsilon, p \nu}^* \Big] \Big[ \left\langle s v^* \right\rangle \rho_{\zeta, p \nu} \rho_{\varepsilon, s \nu}^* + \left\langle v s^* \right\rangle \rho_{\varepsilon, p \nu} \rho_{\zeta, s \nu}^* \Big] \nonumber\\
&- \Big[ \left\langle sv^* \right\rangle \rho_{\zeta, p \nu} \rho_{\varepsilon, s \nu}^* + \left\langle v s^*\right\rangle \rho_{\varepsilon, p \nu} \rho_{\zeta, s \nu}^* \Big] \Big[ \left\langle s v^* \right\rangle \rho_{\zeta, s \nu} \rho_{\varepsilon, p \nu}^* + \left\langle v s^* \right\rangle \rho_{\varepsilon, s \nu} \rho_{\zeta, p \nu}^* \Big] \: , \label{eq:detJ}\\
\mathrm{Tr} \Vie{J}{}{(\nu)} = \: &\mathrm{Tr} \Vie{J}{\mathrm{unco}}{(\nu)} + 2 \Re \Big[ \left\langle sv^* \right\rangle \big( \rho_{\zeta, p \nu} \rho_{\varepsilon, p \nu}^* + \rho_{\zeta, s \nu} \rho_{\varepsilon, s \nu}^* \big) \Big] \: .\label{eq:TrJ}
\end{align}
\label{eq:DOP:general}
\end{subequations}
\end{widetext}
%

\section{Conditions for depolarization}
\label{sec:depolarization}

Equations~\eqref{eq:DOP} and \eqref{eq:DOP:general} provide a general expression for the degree of polarization for a weakly disordered layer in the single scattering regime. This expression allows us to analyze the conditions for depolarization of the scattered light, given an incident polarized plane wave. 
From the properties (\ref{eq:pola:properties}) of the polarization coupling factors, we easily find that for normal incidence ($\Vie{p}{0}{}=\Vie{0}{}{}$), and independently of the incident polarization, one has $\Cie{P}{}{(\nu)}(\Vie{p}{}{},\Vie{0}{}{}) = 1$, meaning that the scattered waves remain perfectly polarized. This can be seen by noticing that the polarization coupling factors in Eq.~(\ref{eq:DOP:general}) are equal in this case, thus canceling $\mathrm{det} \Vie{J}{}{(\nu)}$. We also find that $\Cie{P}{}{(s)}(\Vie{p}{}{},\Vie{p}{0}{}) = 1$, meaning that no depolarization occurs for an $s$-polarized incident wave. Indeed, for an incident $s$-polarized wave, the field scattered by the surface and the field scattered by the volume are produced in the same polarization state, for all realizations of the scattering medium. Thus, independently of the scattering amplitudes $s(\Vie{p}{}{},\Vie{p}{0}{})$ and $v(\Vie{p}{}{},\Vie{p}{0}{})$, the total scattered field is always perfectly polarized. These two results lead to the conclusion that depolarization in the single scattering regime can only occur for a $p$-polarized incident wave at oblique incidence.

For such a wave, it is also interesting to note that substantial depolarization in the single scattering regime can only be observed for two scattering processes (surface and volume) with similar strengths. Indeed, if one of the processes dominates over the other, then the degree of polarization tends to unity. Consider, for example, the extreme case $\Delta \varepsilon = 0$ and $\zeta \neq 0$ where surface scattering dominates. In this case the scattering amplitude $v(\Vie{p}{}{},\Vie{p}{0}{})$ vanishes, so that $\mathrm{det} \Vie{J}{}{(p)} = 0$ and $\Cie{P}{}{(p)}= 1$, as can be seen from Eqs.~\eqref{eq:DOP} and \eqref{eq:DOP:general}. The same analysis holds for $\zeta = 0$ and $\Delta \varepsilon \neq 0$ where volume scattering dominates. This analysis is consistent with the well-known fact that for a either surface or volume scattering, there is no depolarization in the single scattering regime. Conversely, when surface and volume scattering occur simultaneously, the scattered field is the sum of a field scattered by the surface and a field scattered by the volume weighted by random factors (the scattering amplitudes $s$ and $v$). The resulting polarization state is stochastic, and the degree of polarization can decrease. Qualitatively, to observe substantial depolarization, we can deduce that the two scattering processes (surface and volume) must have different polarization responses and comparable strengths.

\section{Connecting the degree of polarization to surface-volume correlations}
\label{sec:results}

Having these considerations in mind, we will focus on the case of a $p$-polarized wave at oblique incidence interacting with a layer with surface and volume disorders with equal strengths, meaning that $|\varepsilon_2 - \varepsilon_1| \sigma_\zeta = \sigma_\varepsilon L$, and equal correlation lengths $\ell_\zeta = \ell_\varepsilon$. Under theses conditions, the correlation functions of surface and volume disorders are identical, $W_\zeta = W_\varepsilon$, and the scattering amplitudes have equal average intensities $\langle |s|^2 \rangle = \langle |v|^2 \rangle$.  
To analyze the influence of surface-volume correlations on depolarization, it is useful to recast the degree of polarization in the form (see Appendix~\ref{app:correlated})
\begin{widetext}
\begin{equation}
\Cie{P}{}{(p)}(\Vie{p}{}{},\Vie{p}{0}{}) = \left[ 1 -  \frac{4 \rho_{sp}^2 (\rho_{\zeta,pp} - \rho_{\varepsilon,pp})^2 \big(1 - |\gamma(\Vie{p}{}{}-\Vie{p}{0}{})|^2 \big)}{\Big[ \rho_{\zeta, pp}^2 + \rho_{\varepsilon,pp}^2 + 2 \rho_{sp}^2 + 2 \mathrm{Re}\big( \gamma(\Vie{p}{}{}-\Vie{p}{0}{}) \big) \big( \rho_{\zeta,pp} \rho_{\varepsilon,pp} + \rho_{sp}^2 \big) \Big]^2} \right]^{1/2} \! , \label{eq:DOP:final}
\end{equation}
where $\gamma(\Vie{p}{}{})$ is the spectral correlation modulator defined in Eq.~\eqref{eq:cross:spectrum}.
\end{widetext}

\subsection{Vanishing or perfect correlation \label{sect:vanishing_polar}}

\begin{figure*}[ht]
\begin{center}
\includegraphics[width=0.32\textwidth, trim = .6cm 0.4cm .6cm 0.4cm,clip]{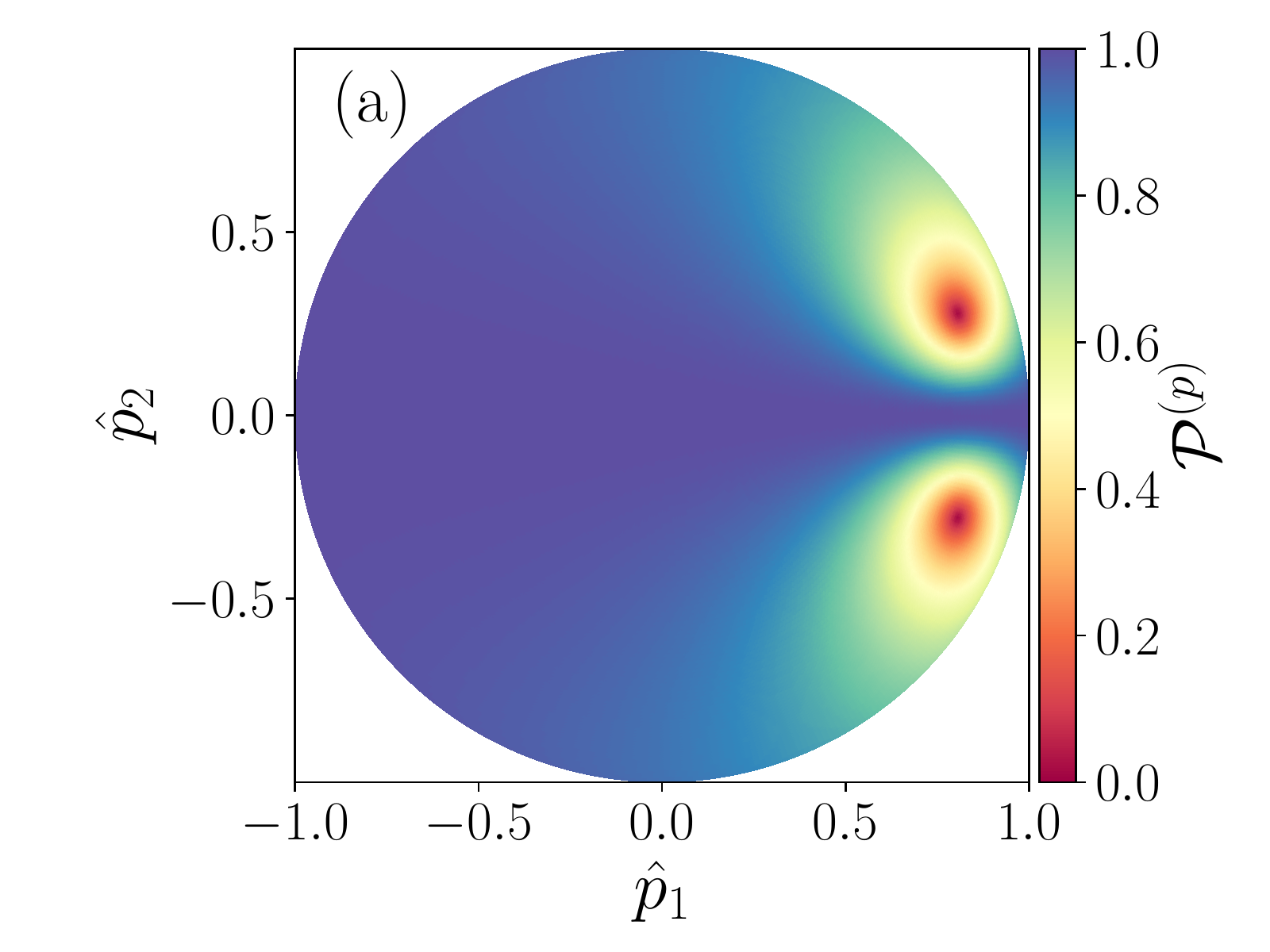}
~
\includegraphics[width=0.32\textwidth, trim = .6cm 0.4cm .6cm 0.4cm,clip]{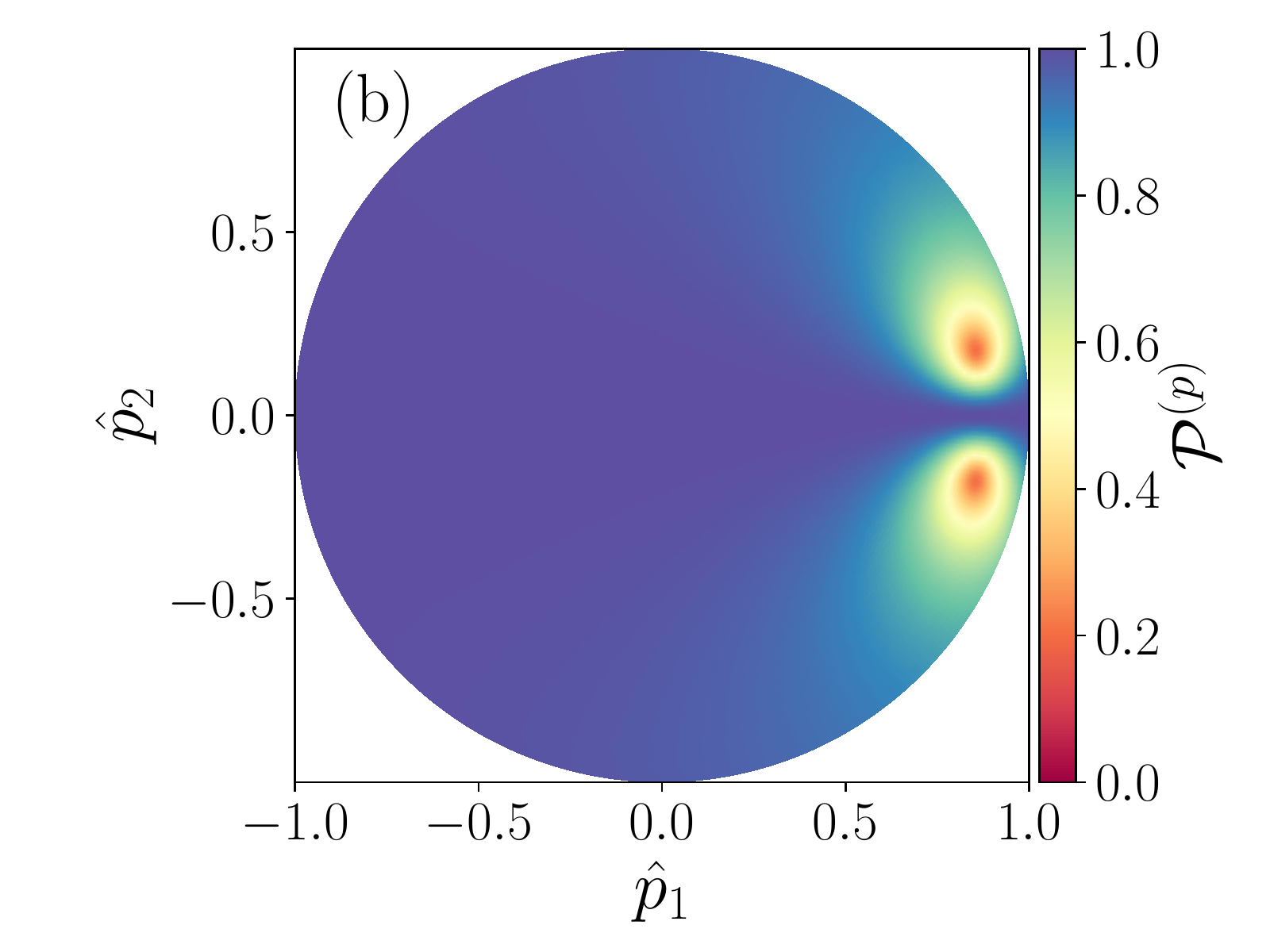}
~
\includegraphics[width=0.32\textwidth, trim = .6cm 0.4cm .6cm 0.4cm,clip]{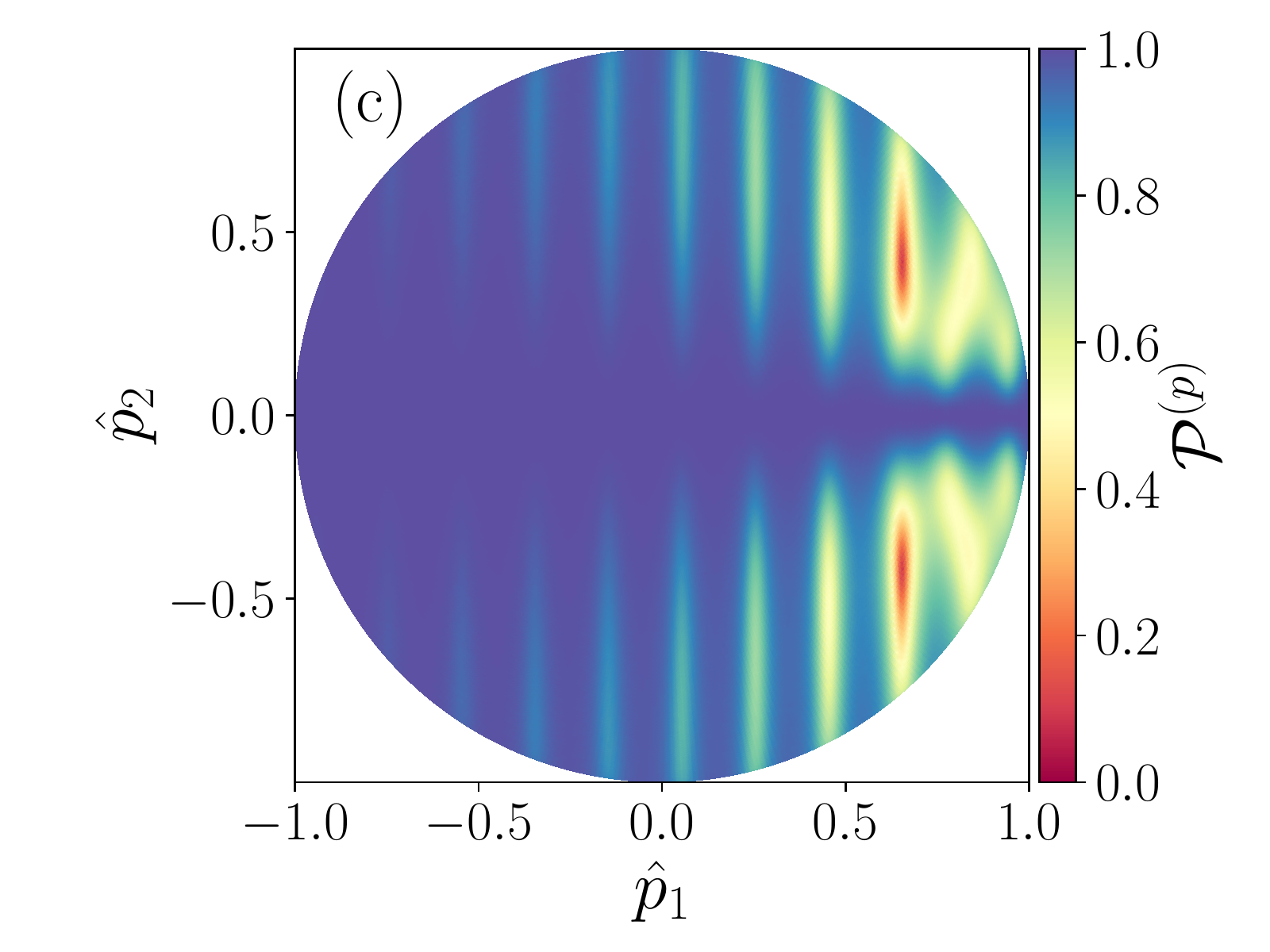}

\caption{ Degree of polarization $\mathcal{P}^{(p)}$ versus the observation direction $\Vie{\hat{p}}{}{} = (\hat{p}_1,\hat{p}_2)$. (a) Uncorrelated surface and volume disorders. (b) Uniform correlation with $\gamma(\Vie{p}{}{} ) = 1/2$. (c) Shift correlation with $\gamma(\Vie{p}{}{}) = \exp \big( i 5 \hat{p}_1 \big) / 2$.  For all cases the angles of incidence are $\theta_0 = 75^\circ$ and $\phi_0 = 0^\circ$. Layer parameters: $\varepsilon_1 = 1$, $\varepsilon_2 = 2.25$, $\sigma_\varepsilon = 0.36$, $L =\lambda / 20$, $\ell_{\varepsilon} = \ell_{\zeta} = \lambda / 2$, $\sigma_\zeta = 1.4 \times 10^{-2} \lambda$. The parameters are chosen such that $\left\langle |s|^2 \right\rangle = \left\langle |v|^2 \right\rangle$.}
\label{fig:dop}
\end{center}
\end{figure*}

The particular case of uncorrelated surface and volume disorders, corresponding to $\gamma(\Vie{p}{}{})=0$, has been examined in detail in Ref.~\onlinecite{OL:Banon:2020}.  It was shown that perfect depolarization can be achieved in specific scattering directions. From Eq.~\eqref{eq:DOP:final}, one immediately finds that for $\gamma(\Vie{p}{}{})=0$ the degree of polarization vanishes when the equality
\begin{equation}
4 \, \rho_{sp}^2 \, \Big( \rho_{\zeta, pp} - \rho_{\varepsilon, pp} \Big)^2 = \Big[ \rho_{\zeta,pp}^2 + \rho_{\varepsilon,pp}^2 + 2 \rho_{sp}^2  \Big]^2
\label{eq:condition2}
\end{equation}
is satisfied. The remarkable fact is that observation directions such that $\rho_{sp} = \rho_{\zeta,pp}= -\rho_{\varepsilon,pp}$ exist, for which condition \eqref{eq:condition2} holds~\cite{OL:Banon:2020}. An example of the angular distribution of the degree of polarization for uncorrelated disorders is presented in Fig.~\ref{fig:dop}(a).  We observe perfect depolarization ($\Cie{P}{}{(p)}=0$) for two scattering directions, symmetrically positioned with respect to the plane of incidence, for which Eq.~\eqref{eq:condition2} is satisfied. In these directions, the field scattered by the surface and the field scattered by the volume are orthogonal and weighted by uncorrelated amplitudes $s$ and $v$ with equal average intensities, leading to perfect depolarization~\cite{OL:Banon:2020}.
\begin{figure*}[ht]
\begin{center}
\includegraphics[width=0.3\textwidth, trim = 3.7cm 1.7cm 3.7cm 1.2cm,clip]{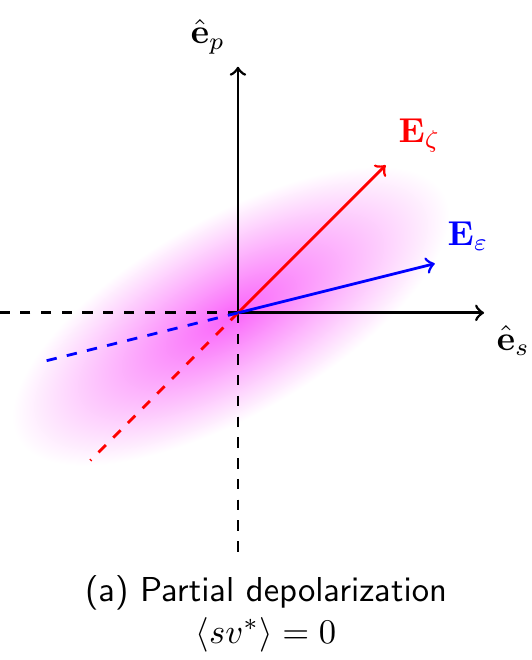}
~
\includegraphics[width=0.3\textwidth, trim = 3.7cm 1.7cm 3.7cm 1.2cm,clip]{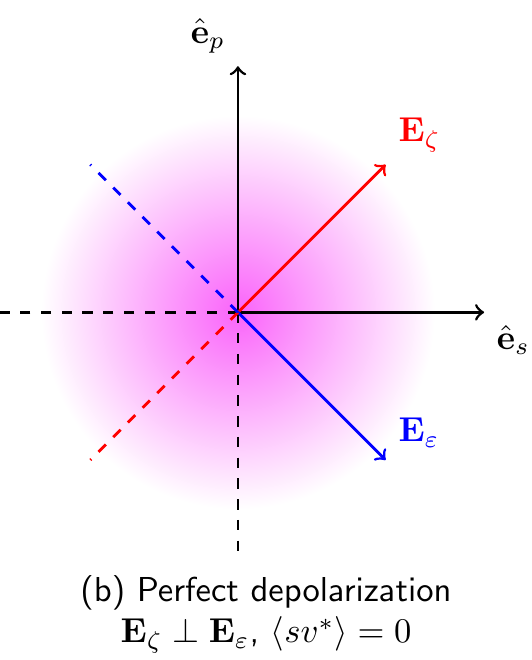}
~
\includegraphics[width=0.3\textwidth, trim = 3.7cm 1.7cm 3.7cm 1.2cm,clip]{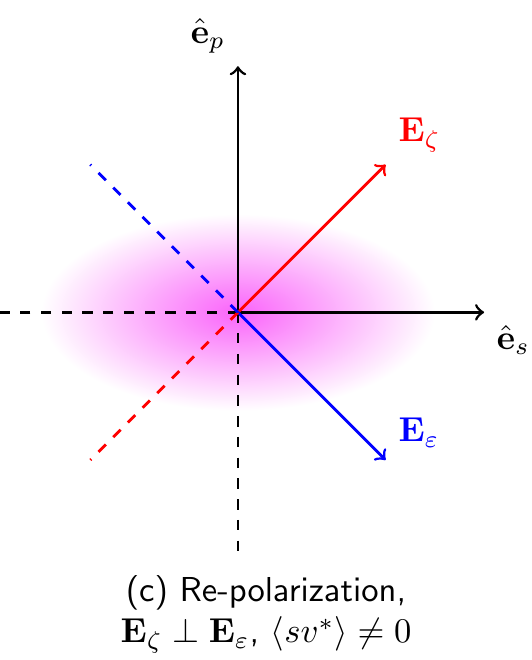}
\caption{Schematic illustration of the conditions for (a) partial depolarization or (b) perfect depolarization of the light scattered from uncorrelated surface and volume disorders. Partial correlation can be seen as a repolarization mechanism, as illustrated in (c). The red and blue arrows represent the deterministic polarization states in the $(\Vie{\hat{e}}{s}{}, \Vie{\hat{e}}{p}{})$ basis. Summing these polarization states with random weights $s$ and $v$ produces the total field whose distribution is represented by the colored area.}
\label{fig:schema}
\end{center}
\end{figure*}

Another extreme situation is that of perfect surface-volume correlation, corresponding to $|\gamma| = 1$. In this case, we find that $\mathcal{P}^{(p)}=1$, independently of the behavior of the polarization coupling factors. This means that single scattering from two perfectly correlated random processes does not induce any depolarization. Indeed, when the two scattering processes are perfectly correlated, the scattering amplitudes $s$ and $v$ are connected by a simple (complex-valued) multiplicative constant. Consequently, even though the polarization states for surface and volume scattering are expected to be different, the resulting scattered field possesses a deterministic polarization state, hence a unit degree of polarization.

\begin{figure*}[ht]
\begin{center}
\includegraphics[width=0.4\textwidth, trim = .6cm 0.4cm 0.5cm 0.4cm,clip]{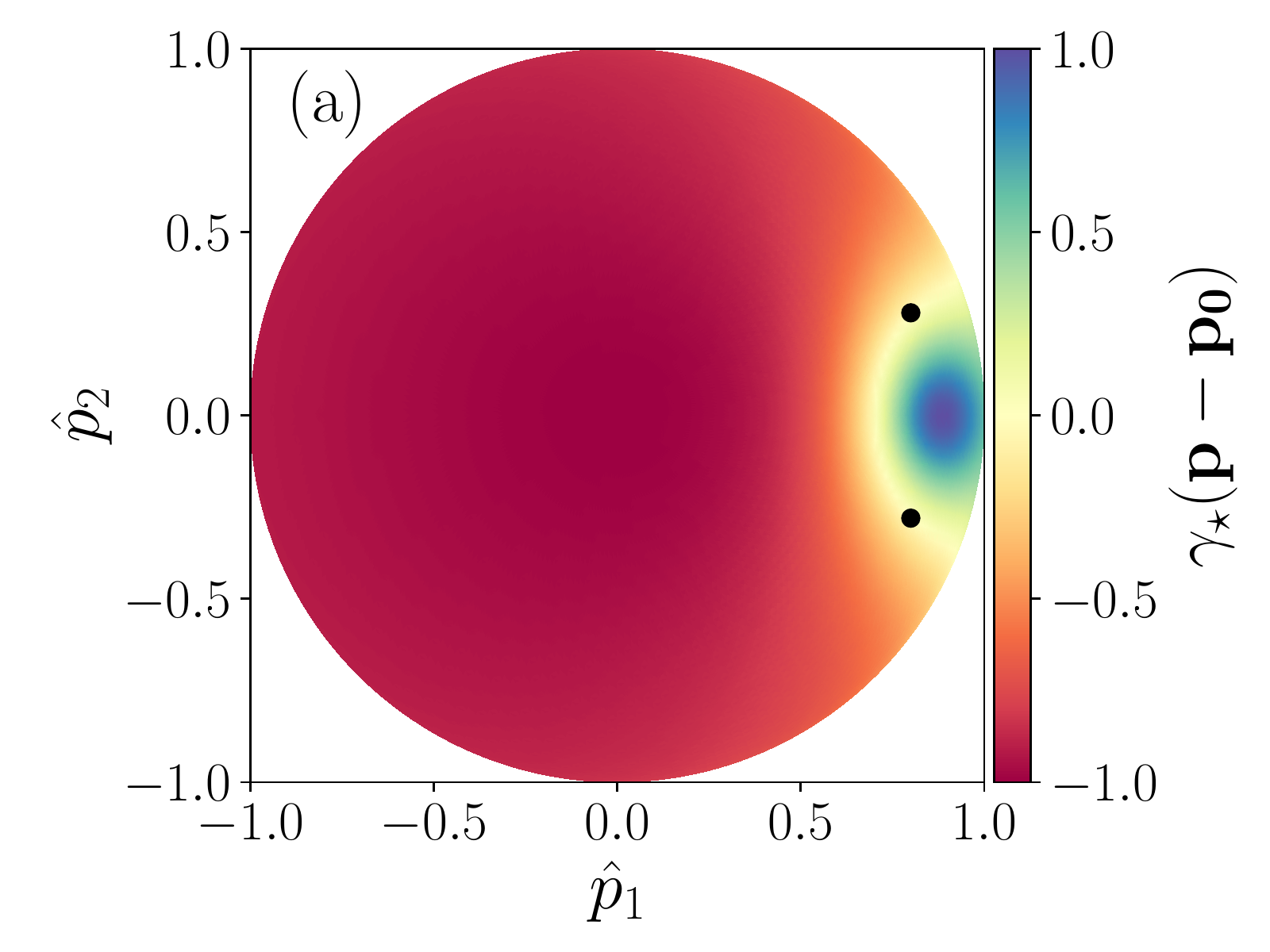}
~
\includegraphics[width=0.4\textwidth, trim = .6cm 0.4cm 0.5cm 0.4cm,clip]{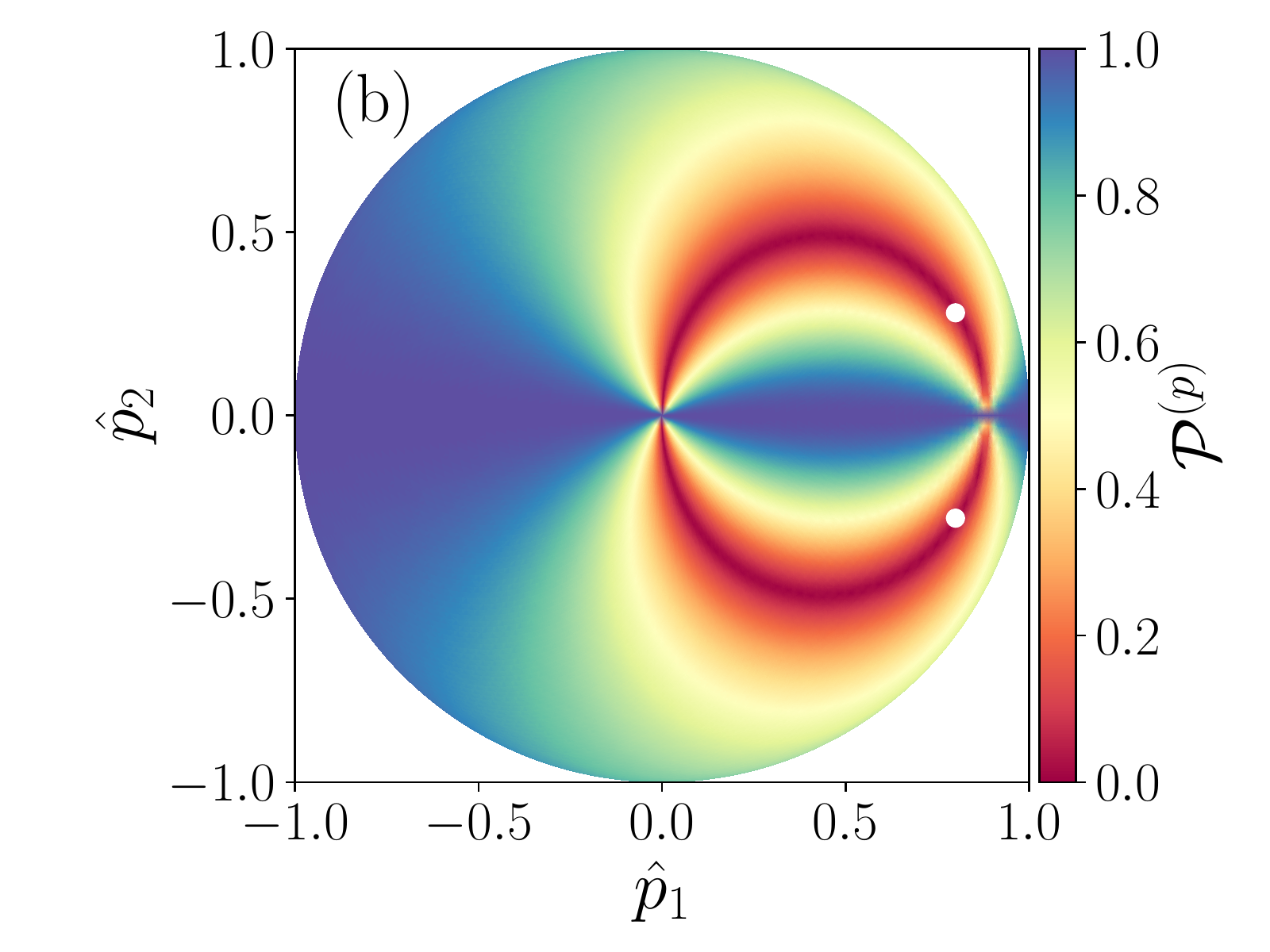}
\caption{ (a) Optimal correlation modulator $\gamma_\star$ versus the observation direction $\Vie{\hat{p}}{}{} = (\hat{p}_1,\hat{p}_2)$ and (b) the corresponding degree of polarization $\Cie{P}{\star}{(p)}(\Vie{p}{}{})$. The black or white dots indicate the directions of perfect depolarization for $\gamma=0$ taken from Fig.~\ref{fig:dop}(a). The remaining parameters are the same as in Fig.~\ref{fig:dop}.}
\label{fig:dop:optimal}
\end{center}
\end{figure*}

\subsection{Partial correlation}

In the presence of partial correlation between the surface and volume disorders, we expect partial depolarization of the scattered light. The direct connection between the degree of polarization and the spectral correlation modulator given by Eq.~\eqref{eq:DOP:final} allows us to study the process quantitatively. In the following we examine a few situations of particular interest. \\

\emph{Uniform correlation} --- A uniform partial correlation is characterized by $\gamma(\Vie{p}{}{}) = \gamma_0$, with $|\gamma_0| < 1$.  Figure~\ref{fig:dop}(b) shows the angular distribution of the degree of polarization for $\gamma_0 = 1/2$. By comparison to Fig.~\ref{fig:dop}(a), we observe that the directions of perfect depolarization for the uncorrelated case correspond to local minima (but not zeros) of the degree of polarization. Moreover, the positions of the minima are shifted compared to the positions of the zeros (with a shift towards larger or smaller azimuthal angles, depending on the sign of $\gamma_0$). \\

\emph{Shift correlation} --- Another particular case is the wave vector dependent correlation modulator $\gamma(\Vie{p}{}{}) = \gamma_0 \exp (i \Vie{a}{}{} \cdot \Vie{p}{}{})$ with $\Vie{a}{}{}$ a constant (spatial) vector. In real space, this form of correlation corresponds to a surface profile and dielectric fluctuations that are scaled and shifted copies of each other, such that $\sigma_\varepsilon \zeta(\Vie{x}{\parallel}{}-\Vie{a}{}{}) = \pm \sigma_\zeta \Delta \varepsilon(\Vie{x}{\parallel}{})$ (for $\gamma_0=\pm 1$).  
 We show in Fig.~\ref{fig:dop}(c) the angular distribution of the degree of polarization for the shift correlation with $\Vie{a}{}{} = 5 \lambda \Vie{\hat{e}}{1}{}$ and $\gamma_0 = 1/2$. Since $|\gamma(\Vie{p}{}{})| = 1/2$ as for the uniform correlation examined previously, the role played by the phase term in $\gamma(\Vie{p}{}{})$ is directly revealed by comparison with Fig.~\ref{fig:dop}(b). We observe in Fig.~\ref{fig:dop}(c) partial depolarization fringes whose positions are controlled by the real part of $\gamma(\Vie{p}{}{})$. These fringes in the degree of polarization are reminiscent of similar fringes observed in the angular distribution of the diffuse intensity~\cite{Banon:2020-1}. \\

\emph{Qualitative picture for partial depolarization} --- We have seen that for perfectly correlated disorders ($|\gamma|=1$), the scattered field possesses a well-defined polarization state, leading to $\mathcal{P}^{(p)}=1$. Conversely, for uncorrelated surface and volume disorders, the fields scattered by the surface and the volume are non-colinear and weighted by random uncorrelated amplitudes $s$ and $v$. The resulting field is partially polarized, as illustrated schematically in Fig.~\ref{fig:schema}(a), and even fully depolarized when the conditions illustrated in Fig.~\ref{fig:schema}(b) are met. Starting from a vanishing degree of polarization for uncorrelated processes, increasing the surface-volume correlation can be seen as a repolarization mechanism. Indeed, even a partial correlation links the weighting amplitudes $s$ and $v$. Consequently, even for orthogonal surface and volume polarization states, the distribution of the resulting field becomes anisotropic as illustrated in Fig.~\ref{fig:schema}(c), leading to partial repolarization for $|\gamma|<1$ and even total repolarization for $|\gamma|=1$. \\

\emph{Engineering the degree of polarization} --- It is also instructive to examine the possibility of shaping the degree of polarization $\Cie{P}{}{(p)}$ by an appropriate design of the surface-volume statistical correlation. For instance, one could seek to cancel the degree of polarization over a range of observation directions, or to set it to some prescribed value. We first note that $\Cie{P}{}{(p)}=1$ in the plane of incidence (since $\rho_{sp} = 0$) independently of the surface-volume correlation, so that shaping is meaningful only for observation directions outside the plane of incidence. For a given direction of incidence $\Vie{p}{0}{}$, consider the problem of minimizing the degree of polarization given by Eq.~(\ref{eq:DOP:final}) in an observation direction $\mathbf{p}$, with $\gamma$ as the free parameter. We note that the minimizer is necessarily real and satisfies $|\gamma|<1$. Indeed, by writing $\gamma = |\gamma| \exp(i \phi)$ and assuming a fixed modulus $|\gamma|$, minimizing Eq.~(\ref{eq:DOP:final}) is equivalent to minimizing 
\begin{align}
 \mathrm{Re}\big( \gamma(\Vie{p}{}{}-\Vie{p}{0}{}) \big) \big( \rho_{\zeta,pp} \rho_{\varepsilon,pp} + \rho_{sp}^2 \big) = \nonumber\\
 |\gamma(\Vie{p}{}{}-\Vie{p}{0}{}) | \cos \phi \big( \rho_{\zeta,pp} \rho_{\varepsilon,pp} + \rho_{sp}^2 \big) \: .
\end{align}
Depending on the sign of the factor $\rho_{\zeta,pp} \rho_{\varepsilon,pp} + \rho_{sp}^2$, we find that the minimum is reached for $\phi=0$ or $\phi=\pi$, forcing $\gamma$ to be real. The problem of minimizing $\Cie{P}{}{(p)}$ is thus reduced to a one-dimensional problem with variable $\gamma$ and is analyzed in detail in Appendix~\ref{App:minimization}. The minimum is found for an optimal correlation modulator $\gamma_\star$ given by 
\begin{equation}
\gamma_\star (\Vie{p}{}{}-\Vie{p}{0}{}) =  - \frac{2 \left( \rho_{\zeta,pp} \rho_{\varepsilon,pp} + \rho_{sp}^2 \right)}{\rho_{\zeta, pp}^2 + \rho_{\varepsilon,pp}^2 + 2 \rho_{sp}^2} \: , \label{eq:gamma:star:text}
\end{equation}
and the corresponding minimum value of the degree of polarization is 
\begin{equation}
\Cie{P}{\star}{(p)}(\Vie{p}{}{}, \Vie{p}{0}{}) = \frac{\left| \rho_{\zeta, pp} + \rho_{\varepsilon,pp} \right|}{ \sqrt{\left( \rho_{\zeta, pp} + \rho_{\varepsilon,pp} \right)^2 + 4 \rho_{sp}^2 } } \: . \label{eq:DOP:mini}
\end{equation}
Note that these expressions are consistent with the existence of directions exhibiting perfect depolarization for uncorrelated disorders.  Indeed, in the absence of surface-volume correlation, the directions $\Vie{p}{}{}$ of perfect depolarization are characterized by the equation $\rho_{\zeta,pp} \rho_{\varepsilon,pp} + \rho_{sp}^2 = 0$~\cite{OL:Banon:2020}.  When this condition is satisfied, we immediately find from Eq.~(\ref{eq:gamma:star:text}) that $\gamma_\star = 0$. Since in these directions we also have $\rho_{\zeta,pp}= -\rho_{\varepsilon,pp}$ (see section~\ref{sect:vanishing_polar}), we also find that the corresponding degree of polarization vanishes.

To get a more general picture in the presence of surface-volume correlations, we show the optimal correlation modulator $\gamma_\star$ versus the observation direction in Fig.~\ref{fig:dop:optimal}(a), and the corresponding degree of polarization in Fig.~\ref{fig:dop:optimal}(b). The two directions corresponding to perfect depolarization for uncorrelated disorders are indicated by the black or white dots in Fig.~\ref{fig:dop:optimal}, both lying on the contours $\gamma_\star = 0$ and $\Cie{P}{\star}{(p)} = 0$ as expected. We also see in Fig.~\ref{fig:dop:optimal}(b) that the degree of polarization vanishes on a contour indicated by the dark red color, and defining a range of observation angles over which the scattered light is fully depolarized. To define this contour, we can set the right-hand side of Eq.~\eqref{eq:DOP:mini} to zero, which leads to the condition $\rho_{\zeta, pp} + \rho_{\varepsilon,pp} = 0$. Recalling the definition of the polarization coupling factors in Eq.~\eqref{eq:rho}, we find that perfect depolarization is obtained for observation directions $\Vie{p}{}{}$ satisfying
\begin{align}
 t_{12}^{(p)}(\Vie{p}{}{}) \Vie{\hat{e}}{2,p}{+}(\Vie{p}{}{}) \cdot \Big[ &\Vie{\hat{e}}{1,p}{-}(\Vie{p}{0}{}) + r_{21}^{(p)}(\Vie{p}{0}{}) \Vie{\hat{e}}{1,p}{+}(\Vie{p}{0}{}) \nonumber\\
 &+ \Vie{\hat{e}}{2,p}{-}(\Vie{p}{0}{}) \, t_{21}^{(p)}(\Vie{p}{0}{})\Big] = 0 \: .
\end{align}
This implicit equation defines the dark red contour in Fig.~\ref{fig:dop:optimal}(b).

\section{Conclusion}
\label{sect:conclusion}

In summary, we have derived a general expression for the degree of polarization of the light scattered from a weakly scattering layer exhibiting both surface and volume scattering. This expression shows the direct connection between the degree of polarization and the cross-correlation function of the surface and volume disorders. We have analyzed depolarization of the backscattered light for uncorrelated, perfectly correlated and partially correlated disorders. The analysis shows that measuring the degree of polarization could be used, in principle, to assess the statistical correlation between the surface roughness and the bulk dielectric fluctuations. In addition, an appropriate shaping of the correlation function could be used to shape the degree of polarization over a range of scattering angles.  The latter could be more easily achieved with thin films with correlated surfaces (for which the general analysis developed in this work also holds) by successive and controlled exposure of speckle patterns onto photosensitive coatings.

\section*{Funding}
LABEX WIFI (Laboratory of Excellence within the French Program Investments for the Future) under references ANR-10- LABX-24 and ANR-10-IDEX-0001-02 PSL*.  The French National Research Agency under the contract ANR-15-CHIN-0003. 

\section*{Acknowledgments}

We are grateful to Eugenio M\'{e}ndez for fruitful discussions.

\section*{Disclosures}

The authors declare no conflicts of interest.

\section*{Data Availability}

Numerical data underlying the results presented in this paper were produced using a homemade computer program. They are not publicly available at this time but may be obtained from the authors upon reasonable request.

\newpage

\begin{widetext}

\appendix

\section{Expression for the degree of polarization for a p-polarized incident wave}\label{app:correlated}

In this appendix we derive Eq.~(\ref{eq:DOP:final}). For correlated surface and volume disorders, making use of properties (\ref{eq:pola:properties}), it is easy to show that Eq.~(\ref{eq:detJ}) can be rewritten as
\begin{align}
\mathrm{det} \Vie{J}{}{(p)} = &\left\langle |s|^2 \right\rangle \left\langle |v|^2 \right\rangle \left| \rho_{\zeta, p p}  - \rho_{\varepsilon, p p} \right|^2  \rho_{s p}^2 + 4 \mathrm{Re} \big( \left\langle s v^* \right\rangle \big)^2 \rho_{\zeta,pp} \, \rho_{\varepsilon, pp} \, \rho_{sp}^2  - \Big| \rho_{\zeta,pp} \left\langle s v^* \right\rangle + \rho_{\varepsilon, pp} \left\langle v s^* \right\rangle  \Big|^2 \rho_{sp}^2 \: .
\end{align}
From the expressions for the various covariances derived in Ref.~\onlinecite{Banon:2020-1} (see Eqs.~(D4-D6) in~\cite{Banon:2020-1})  we obtain
\begin{align}
\mathrm{det} \Vie{J}{}{(p)} (\Vie{p}{}{},\Vie{p}{0}{}) = &\left[ \frac{k_0^4}{4 |\alpha_2(\Vie{p}{}{})|^2} \right]^2 (\varepsilon_2 -\varepsilon_1)^2 \sigma_\zeta^2 \sigma_\varepsilon^2 L^2 \widetilde{W}_\zeta (\Vie{p}{}{} - \Vie{p}{0}{}) \widetilde{W}_\varepsilon (\Vie{p}{}{} - \Vie{p}{0}{}) \, \rho_{s p}^2(\Vie{p}{}{},\Vie{p}{0}{}) \nonumber\\
&\times \Bigg[ \left| \rho_{\zeta, p p}(\Vie{p}{}{},\Vie{p}{0}{})  - \rho_{\varepsilon, p p}(\Vie{p}{}{},\Vie{p}{0}{}) \right|^2   + 4 \mathrm{Re} ( \gamma(\Vie{p}{}{}-\Vie{p}{0}{}) )^2 \rho_{\zeta,pp}(\Vie{p}{}{},\Vie{p}{0}{}) \, \rho_{\varepsilon, pp}(\Vie{p}{}{},\Vie{p}{0}{})  \nonumber\\
&- \Big| \rho_{\zeta,pp}(\Vie{p}{}{},\Vie{p}{0}{}) \gamma(\Vie{p}{}{}-\Vie{p}{0}{}) + \rho_{\varepsilon, pp}(\Vie{p}{}{}-\Vie{p}{0}{}) \gamma^*(\Vie{p}{}{}-\Vie{p}{0}{})  \Big|^2 \Bigg] \nonumber \\
= &\left[ \frac{k_0^4}{4 |\alpha_2(\Vie{p}{}{})|^2} \right]^2 (\varepsilon_2 -\varepsilon_1)^2 \sigma_\zeta^2 \sigma_\varepsilon^2 L^2 \widetilde{W}_\zeta (\Vie{p}{}{} - \Vie{p}{0}{}) \widetilde{W}_\varepsilon (\Vie{p}{}{} - \Vie{p}{0}{}) \, \rho_{s p}^2(\Vie{p}{}{},\Vie{p}{0}{}) \nonumber\\
&\times \Bigg[ \left| \rho_{\zeta, p p}(\Vie{p}{}{},\Vie{p}{0}{})  - \rho_{\varepsilon, p p}(\Vie{p}{}{},\Vie{p}{0}{}) \right|^2  \nonumber\\
&+ \rho_{\zeta, p p}(\Vie{p}{}{},\Vie{p}{0}{}) \rho_{\varepsilon, p p}(\Vie{p}{}{},\Vie{p}{0}{}) \Big( 4 \mathrm{Re} ( \gamma(\Vie{p}{}{}-\Vie{p}{0}{}) )^2 - \gamma^2(\Vie{p}{}{}-\Vie{p}{0}{}) -\gamma^{*2}(\Vie{p}{}{}-\Vie{p}{0}{}) \Big) \nonumber \\
&- |\gamma(\Vie{p}{}{}-\Vie{p}{0}{})|^2 \Big( \rho_{\zeta,pp}^2(\Vie{p}{}{},\Vie{p}{0}{}) + \rho_{\varepsilon, pp}^2(\Vie{p}{}{},\Vie{p}{0}{}) \Big) \Bigg] \: .
\end{align}
Making use of the identity $4 (\mathrm{Re} \, z)^2 - z^2 -z^{*2} = 2 |z|^2$ valid for any complex number $z$, we finally obtain
\begin{align}
\mathrm{det} \Vie{J}{}{(p)} (\Vie{p}{}{},\Vie{p}{0}{})
= \: &\left[ \frac{k_0^4}{4 |\alpha_2(\Vie{p}{}{})|^2} \right]^2 (\varepsilon_2 -\varepsilon_1)^2 \sigma_\zeta^2 \sigma_\varepsilon^2 L^2 \widetilde{W}_\zeta (\Vie{p}{}{} - \Vie{p}{0}{}) \widetilde{W}_\varepsilon (\Vie{p}{}{} - \Vie{p}{0}{}) \, \rho_{s p}^2(\Vie{p}{}{},\Vie{p}{0}{}) \nonumber\\
&\times \left| \rho_{\zeta, p p}(\Vie{p}{}{},\Vie{p}{0}{})  - \rho_{\varepsilon, p p}(\Vie{p}{}{},\Vie{p}{0}{}) \right|^2  \Big( 1 - |\gamma(\Vie{p}{}{}-\Vie{p}{0}{})|^2 \Big) \nonumber\\
= \:  &\mathrm{det} \Vie{J}{\mathrm{uncor}}{(p)} (\Vie{p}{}{},\Vie{p}{0}{}) \, \Big( 1 - |\gamma(\Vie{p}{}{}-\Vie{p}{0}{})|^2 \Big) \: . \label{eq:detJcorr}
\end{align}
To complete the derivation of the degree of polarization, we need to compute the trace of the coherence matrix, and we obtain
\begin{align}
\mathrm{Tr} \Vie{J}{}{(p)} (\Vie{p}{}{},\Vie{p}{0}{})  = &\frac{k_0^4}{4 |\alpha_2(\Vie{p}{}{})|^2} \Bigg[ (\varepsilon_2 -\varepsilon_1)^2 \sigma_\zeta^2 \widetilde{W}_\zeta (\Vie{p}{}{}-\Vie{p}{0}{}) \big( \rho_{\zeta,pp}^2 + \rho_{sp}^2 \big) + \sigma_\varepsilon^2 L^2 \widetilde{W}_\varepsilon (\Vie{p}{}{}-\Vie{p}{0}{}) \big( \rho_{\varepsilon,pp}^2 + \rho_{sp}^2 \big) \nonumber\\
&+ 2 \mathrm{Re} \big( \gamma(\Vie{p}{}{}-\Vie{p}{0}{}) \big) (\varepsilon_2 -\varepsilon_1) \sigma_\zeta \widetilde{W}_\zeta^{1/2} (\Vie{p}{}{}-\Vie{p}{0}{}) \sigma_\varepsilon L \widetilde{W}_\varepsilon^{1/2} (\Vie{p}{}{}-\Vie{p}{0}{}) \Big( \rho_{\zeta,pp} \rho_{\varepsilon,pp} + \rho_{sp}^2 \Big) \Bigg] \\
= & \mathrm{Tr} \Vie{J}{\mathrm{unco}}{(p)} (\Vie{p}{}{},\Vie{p}{0}{}) \nonumber\\
&+ 2 \mathrm{Re} \big( \gamma(\Vie{p}{}{}-\Vie{p}{0}{}) \big) (\varepsilon_2 -\varepsilon_1) \sigma_\zeta \widetilde{W}_\zeta^{1/2} (\Vie{p}{}{}-\Vie{p}{0}{}) \sigma_\varepsilon L \widetilde{W}_\varepsilon^{1/2} (\Vie{p}{}{}-\Vie{p}{0}{}) \Big( \rho_{\zeta,pp} \rho_{\varepsilon,pp} + \rho_{sp}^2 \Big) \: . \nonumber
\end{align}
In the conditions $|\varepsilon_2 - \varepsilon_1| \sigma_\zeta = \sigma_\varepsilon L$ and $\ell_\zeta = \ell_\varepsilon$, that are assumed in the main text, we obtain after some simplifications
\begin{equation}
\Cie{P}{}{(p)}(\Vie{p}{}{},\Vie{p}{0}{}) = \left[ 1 -  \frac{4 \rho_{sp}^2 (\rho_{\zeta,pp} - \rho_{\varepsilon,pp})^2 \big(1 - |\gamma(\Vie{p}{}{}-\Vie{p}{0}{})|^2 \big)}{\Big[ \rho_{\zeta, pp}^2 + \rho_{\varepsilon,pp}^2 + 2 \rho_{sp}^2 + 2 \mathrm{Re}\big[ \gamma(\Vie{p}{}{}-\Vie{p}{0}{}) \big] \big( \rho_{\zeta,pp} \rho_{\varepsilon,pp} + \rho_{sp}^2 \big) \Big]^2} \right]^{1/2} \: ,
\end{equation}
which is Eq.~(\ref{eq:DOP:final}) in the main text.

\section{Minimization of the degree of polarization} \label{App:minimization}

In this appendix we derive Eqs.~(\ref{eq:gamma:star:text}) and  (\ref{eq:DOP:mini}). We have seen in section~\ref{sec:results} that the degree of polarization may be minimized with $\gamma$ as a free parameter, and that the minimizer is real valued. This means that we can search the point $|\gamma_\star|$ such that $\partial \mathcal{P}^{(p)} / \partial |\gamma| = 0$, or equivalently, $\partial \left(\mathcal{P}^{(p)} \right)^2/ \partial |\gamma| = 0$. Using the notations $A = 4 \rho_{sp}^2 (\rho_{\zeta,pp} - \rho_{\varepsilon,pp})^2$,  $B = \rho_{\zeta, pp}^2 + \rho_{\varepsilon,pp}^2 + 2 \rho_{sp}^2$, and $C = 2 \big| \rho_{\zeta,pp} \rho_{\varepsilon,pp} + \rho_{sp}^2 \big|$,
we can write
\begin{equation}
\left[ \Cie{P}{}{(p)} \right]^2 = 1 -  \frac{A\big(1 - |\gamma|^2 \big)}{\Big[ B - C |\gamma| \Big]^2}  \: , \label{eq:DOP:2}
\end{equation}
from which we find that
\begin{equation}
\frac{\partial \left[ \Cie{P}{}{(p)} \right]^2}{\partial |\gamma| } = 2A  \times \frac{|\gamma| (B - C |\gamma|) - C(1 - |\gamma|^2) }{\Big[ B - C |\gamma| \Big]^3}  \: . \label{eq:dDOP:2}
\end{equation}
The minimizer $|\gamma_\star|$ is the solution to the equation
\begin{equation}
|\gamma^\star| (B - C |\gamma_\star|) - C(1 - |\gamma_\star|^2)  = 0 \: ,
\end{equation}
which immediately leads to
\begin{equation}
|\gamma_\star| = \frac{C}{B} = \frac{2 \big| \rho_{\zeta,pp} \rho_{\varepsilon,pp} + \rho_{sp}^2 \big|}{\rho_{\zeta, pp}^2 + \rho_{\varepsilon,pp}^2 + 2 \rho_{sp}^2} \: .
\end{equation}
Since $\mathrm{sign}(\gamma_\star) = - \mathrm{sign} ( \rho_{\zeta,pp} \rho_{\varepsilon,pp} + \rho_{sp}^2 )$, we end up with
\begin{equation}
\gamma_\star= - \frac{2 \left( \rho_{\zeta,pp} \rho_{\varepsilon,pp} + \rho_{sp}^2 \right)}{\rho_{\zeta, pp}^2 + \rho_{\varepsilon,pp}^2 + 2 \rho_{sp}^2} \: . \label{eq:gamma:star}
\end{equation}
By inserting Eq.~\eqref{eq:gamma:star} into Eq.~\eqref{eq:DOP:2} we also find that
\begin{align}
\left[ \Cie{P}{\star}{(p)} \right]^2 &= 1 -  \frac{A\big(1 - \frac{C^2}{B^2} \big)}{\Big[ B -  \frac{C^2}{B} \Big]^2} = 1 - \frac{A}{B^2 - C^2} \: , \nonumber\\
&= 1 - \frac{4 \rho_{sp}^2 (\rho_{\zeta,pp} - \rho_{\varepsilon,pp})^2}{ \left( \rho_{\zeta, pp}^2 + \rho_{\varepsilon,pp}^2 + 2 \rho_{sp}^2 \right)^2 - 4 \left( \rho_{\zeta,pp} \rho_{\varepsilon,pp} + \rho_{sp}^2 \right)^2 } \: .
\end{align}
The denominator of the second term on the right-hand side can be recast as
\begin{align}
\left( \rho_{\zeta, pp}^2 + \rho_{\varepsilon,pp}^2 + 2 \rho_{sp}^2 \right)^2 - 4 \left( \rho_{\zeta,pp} \rho_{\varepsilon,pp} + \rho_{sp}^2 \right)^2  =
 \left(   \rho_{\zeta, pp} - \rho_{\varepsilon,pp}\right)^2  \left[  \left( \rho_{\zeta, pp} + \rho_{\varepsilon,pp} \right)^2 + 4 \rho_{sp}^2  \right] \: ,
\end{align}
which finally leads to
\begin{equation}
\left[ \Cie{P}{\star}{(p)} \right]^2 = \frac{\left( \rho_{\zeta, pp} + \rho_{\varepsilon,pp} \right)^2}{ \left( \rho_{\zeta, pp} + \rho_{\varepsilon,pp} \right)^2 + 4 \rho_{sp}^2 } \: . \label{eq:DOP2:mini}
\end{equation}
This completes the derivation of Eqs.~(\ref{eq:gamma:star:text}) and  (\ref{eq:DOP:mini}) in the main text.

\end{widetext}

%
\bibliography{arxiv}

\end{document}